\documentclass[a4paper,11pt]{article}
\pdfoutput=1
\usepackage{jinstpub}
\usepackage{graphicx}
\usepackage{lineno}
\title{\boldmath The IceCube-Gen2 Neutrino Observatory}

\author[a,1]{B. A. Clark,\note{Corresponding author.}}
\affiliation[a]{Dept. of Physics and Astronomy, Michigan State University,\\567 Wilson Rd, East Lansing, MI 48824, USA}
\emailAdd{baclark@msu.edu}

\collaboration[c]{on behalf of the IceCube-Gen2 Collaboration$^*$\note[*]{Full author list and acknowledgments are available at \href{https://icecube.wisc.edu/collaboration/authors/\#collab=IceCube-Gen2&date=2021-05-18&formatting=web&tag=VLVnT+2021}{icecube.wisc.edu}.}}

\abstract{The IceCube Neutrino Observatory opened the window on neutrino astronomy by discovering high-energy astrophysical neutrinos in 2013 and identifying the first compelling astrophysical neutrino source, the blazar TXS0506+056, in 2017. In this talk, we will discuss the science reach and ongoing development of the IceCube-Gen2 facility---a planned extension to IceCube. IceCube-Gen2 will increase the rate of observed cosmic neutrinos by an order of magnitude, be able to detect five-times fainter neutrino sources, and extend the measurement of astrophysical neutrinos several orders of magnitude higher in energy. We will discuss the envisioned design of the instrument, which will include an enlarged in-ice optical array, a surface array for the study of cosmic-rays, and a shallow radio array to detect ultra-high energy ($>100$\,PeV) neutrinos. we will also highlight ongoing efforts to develop and test new instrumentation for IceCube-Gen2.}

\keywords{Neutrino detectors, Large detector systems for particle and astroparticle physics}

\proceeding{9$^{\text{th}}$ Very Large Volume Neutrino Telescope Workshop (VLVnT-2021)\\
  18-21 May 2021\\
  Valencia, Spain}

\notoc
\begin{document}
\maketitle
\flushbottom

\section{Introduction}
High-energy neutrinos are unique messengers to the high-redshift reaches of the Universe. Unlike photons and cosmic-rays, as chargeless and weakly interacting particles, neutrinos travel unattenuated and undeflected across the cosmos. The IceCube Neutrino Telescope detects neutrinos by deploying photomultiplier tubes (PMTs) in Digital Optical Modules (DOMs) deep in the Antarctic glacier, and looking for the Cherenkov light produced by the byproducts of neutrino-matter interactions in the ice.  IceCube discovered a flux of astrophysical neutrinos in 2013~\cite{Aartsen:2013jdh}.
Using various observation methods, IceCube has observed astrophysical neutrinos between 10 TeV and 10 PeV. The spectrum is well described by a power law in energy with index ranging from -2.28 to -2.89 \cite{Stettner:2019tok, Aartsen:2020aqd, Aartsen:2018vez, Abbasi:2020jmh}.
The arrival directions of the neutrinos are consistent with isotropy after accounting for the effects of Earth absorption. It also appears the flux is ``flavor democratic," meaning all three flavors of neutrinos contribute roughly equally to the total flux.
IceCube has started probing the flavor composition of the neutrinos through the identification of the first astrophysical electron anti-neutrino at the Glashow resonance~\cite{IceCube:2021rpz} as well the first tau-neutrinos~\cite{IceCube:2020abv}.
Observation of neutrinos from the direction of the blazar TXS0506+056 in 2017 provided a first convincing candidate for a point source of neutrinos~\cite{IceCube:2018cha, IceCube:2018dnn}. In pursuit of further sources, IceCube regularly announces the detection of neutrinos with a high-probability of astrophysical origin to the community through ``realtime-alerts"~\cite{Blaufuss:2019fgv}, which has led to followup and the identification of other possible sources~\cite{Stein:2020xhk}.

To make further progress, next generation observatories must enable improved precision on the spectrum and composition of the diffuse flux, the identification of more point sources, and better-cross correlation with astronomical catalogs. This requires higher event rates over a broader range in energy, and improved angular resolution. A variety of next generation observatories are being proposed or constructed to meet these challenges, including KM3NeT~\cite{Adrian-Martinez:2016fdl}, Baikal-GVD~\cite{1997APh.....7..263B}, P-ONE~\cite{Agostini:2020aar}, and IceCube-Gen2~\cite{Aartsen:2020fgd}; the latter of these is the topic of these proceedings.

In these proceedings, we will discuss the science reach and ongoing development of the IceCube-Gen2 facility, which is approximately an order of magnitude larger than IceCube. The anticipated and unanticipated successes of the IceCube detector promises that Gen2 will provide unprecedented insight into the spectrum and sources of astrophysical neutrinos. In Sec.~\ref{sec:facility}, we discuss the envisioned design of the instrument. In Sec.~\ref{sec:performance} we describe the expected performance of the detector and the potential science reach. In Sec.~\ref{sec:instrumentation} we describe ongoing work to develop instrument for the detector, and in Sec.~\ref{sec:conclusion} we conclude.

\section{The IceCube-Gen2 Facility}

\label{sec:facility}
\begin{figure}[]
    \centering
    \includegraphics[width=\textwidth]{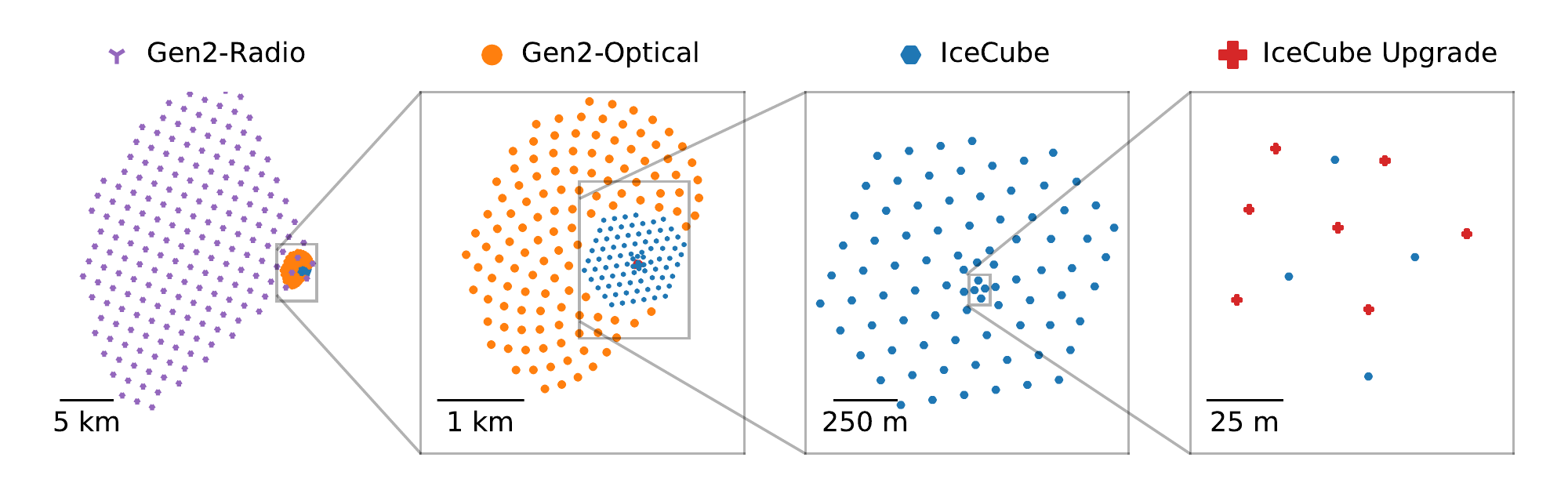}
    \caption{A top-down view of the IceCube-Gen2 facility.}
    \label{fig:gen2_facility}
\end{figure}

The IceCube-Gen2 facility will be a broadband neutrino observatory, targeting physics from GeV to EeV energies. A top-down view of the facility is provided in Fig.~\ref{fig:gen2_facility}. The IceCube-Gen2 observatory builds on the success of the existing IceCube, IceTop, and DeepCore instruments with the addition of four new components:

\begin{enumerate}
    \item \textbf{The IceCube Upgrade}: Visible in red, the IceCube Upgrade is a five-year construction project currently underway~\cite{Ishihara:2019aao}. The Upgrade will consist of 7 new strings with 693 optical modules. Key goals for the Upgrade including enhancing the sensitivity to GeV scale neutrinos as well as improved understanding and calibration of the ice itself~\cite{Rongen:VLVNT2021}. 
    The IceCube upgrade also serves as a platform for developing Gen2 technologies, such as pixelated optical modules and wavelength shifting sensors.
    \item \textbf{Gen2-Optical}: Visible in orange, Gen2-Optical is envisioned as an enlarged optical array for the detection of TeV-PeV neutrinos. The baseline design features 120 new strings laid out in a ``Sunflower" pattern with a 240\,m lateral spacing for a total volume of 8\,km$^3$ the geometry is still being optimized. Each string will contain $\sim$80 optical modules distributed 17\,m apart vertically between 1340-2600\, depth. 
    \item \textbf{Gen2-Surface}: Co-located on top of each new Gen2-Optical string will be a cosmic-ray detecting station. This extension of IceTop will feature two complimentary techonologies: scintillator panels and radio antennas, allowing for separation of the muonic and electromagnetic component of a cosmic-ray air shower. Besides measuring PeV cosmic rays, the surface array also provides a veto to the in-ice array for downgoing cosmic-ray induced air showers.
    \item \textbf{Gen2-Radio}: Visible in purple, the Gen2-Radio array will be $\mathcal{O}(200)$ clusters of antennas deployed over $\sim500\rm{km}^2$ for the detection of EeV neutrinos. The radio stations will be a combination of antennas deployed both at the surface and deeper in the ice at $\sim200$\,m depth, with the final design still being optimized. The main trigger will be a phased-array system which beamforms in realtime to lower triggering thresholds.
\end{enumerate}

\section{Performance}
\label{sec:performance}
\begin{figure}[]
    \centering
    \includegraphics[width=0.49 \textwidth]{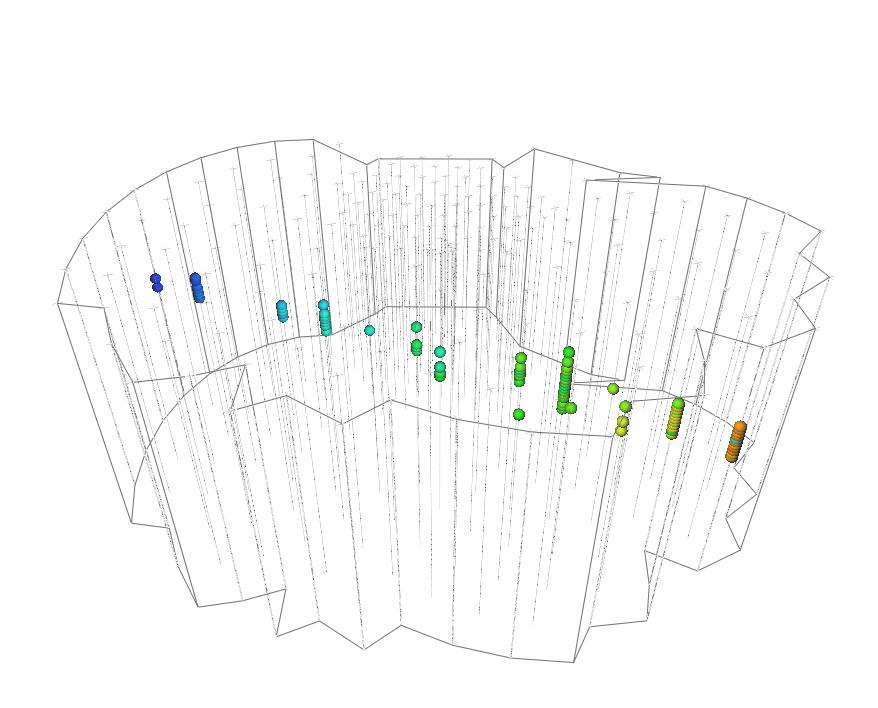}
    \includegraphics[width=0.49 \textwidth]{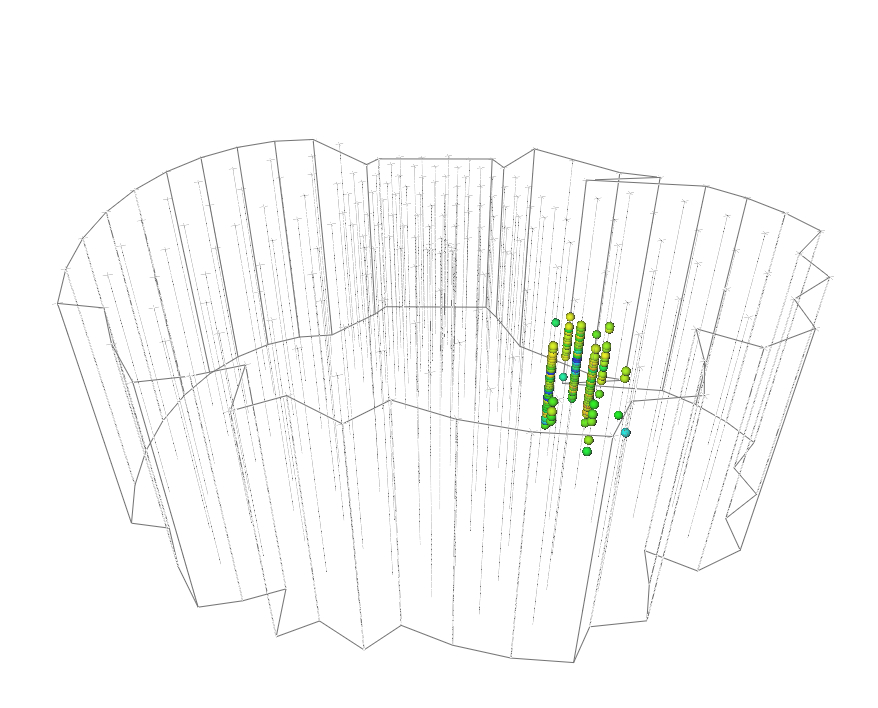}
    \caption{A view of the two main ``topologies" of neutrino interactions expected in the optical component of IceCube-Gen2: ``tracks" (left, a 320 TeV event) ``cascades" (right, a 120 TeV event). The color of a module indicates the relative arrival time of the light (with red indicating earlier arrival times, and blue indicating later arrival times). The size of a module is proportional to the quantity of the light recorded.}
    \label{fig:events}
\end{figure}

In this section we describe some of the key performance metrics for the detector, and highlight some of science reach. For a comprehensive discussion, a full White Paper is available~\cite{Aartsen:2020fgd}. Focusing on the optical component of IceCube-Gen2, neutrino events are expected in two main ``topologies," both of which are visible in Fig.~\ref{fig:events}. Charged-current interactions of muon neutrinos give rise to long-lived daughter muons which can cross the entire length of the detector, giving rise to ``tracks." Neutral-current interactions of all neutrino flavors, and charged-current interactions of electrons and tau neutrinos give rise to roughly spherical depositions of light, or ``cascades". More exotic topologies are possible for very-high energy tau neutrinos, which travel roughly 50\,m/PeV. At high enough energies, the tau can travel macroscopic distances before decaying, resulting in a so-called ``double-bang" signature; at very high energies the tau can even appear as a track.

IceCube-Gen2 will provide a roughly ten-fold increase in effective volume for cascades compared to IceCube, which is critical to measuring super-PeV events and better characterizing the astrophysical flux above 100 TeV.
Using the high-energy starting event (HESE) technique, where the outer layers of the detector are used to reject atmospheric muons \cite{Abbasi:2020jmh}, raises the energy threshold from 60 TeV  in IceCube to 200 TeV in Gen2~\cite{Aartsen:2020fgd}.
The best sensitivity to tracks is achieved with so-called ``through-going" tracks, in which the neutrino interaction vertex is outside the detector.
This class of events has historically provided the best sensitivity in the search for point sources of neutrinos. For high-quality through-going tracks, the effective area is increased by roughly a factor of five relative to IceCube; the median angular error, or pointing resolution, of such events is improved by roughly a factor of two~\cite{Aartsen:2020fgd}.

One of the most challenging backgrounds in the search for high-energy neutrinos is the large flux of atmospheric muons and neutrinos from the Southern Sky which outnumbers the astrophysical neutrino flux by more than nine orders of magnitude. The enlarged Gen2-Surface array will increase the fraction of the sky for which the surface array serves as a veto. The total acceptance for coincidence events which can be vetoed by the surface array is increased by over an order of magnitude relative to IceCube, increasing from 0.25\,km$^2$sr to 10\,km$^2$sr~\cite{Aartsen:2020fgd}.

\section{Science Reach}
With the new, larger array, IceCube-Gen2 will have unprecedented sensitivity to both steady and transient sources of neutrinos, for example blazars/Active Galactic Nuclei, Gamma Ray Bursts, neutron star mergers, etc. For steady sources, IceCube-Gen2 will be sensitive to emitters five times fainter than IceCube; combined with the improved angular resolution, this provides the potential to observe an order of magnitude more individual neutrino sources. The sensitivity to transients is likewise improved. Quantified by the volume of the observable universe from which a transient can be observed, for a transient of 100\,s duration and luminosity of $10^{50}$ ergs, the observable volume increases by an order of magnitude up to 10\,PeV~\cite{Aartsen:2020fgd}.

Many searches for neutrino sources expect the neutrinos to be spatially and temporally coincident with other messengers, for example electromagnetic radiation or gravitational waves. With many next generation instruments coming online in the next decade, e.g. the Vera Rubin Observatory, the Cherenkov Telescope Array, and advanced LIGO, IceCube-Gen2 will have an important role to play in multimessenger astronomy. IceCube-Gen2 will also be sensitive to so-called ``hidden sources" which might reside in optically-thick environments where only the neutrinos escape. For example, the 156 day long neutrino flare from TXS0506+056 would have been observed in IceCube-Gen2 at $>13\sigma$ significance, \textit{without} a coincident detection in gamma rays~\cite{Aartsen:2020fgd}.

IceCube-Gen2 will also probe the spectrum of high-energy neutrinos over an unprecedented range of energies. The optical and radio detectors combined will provide continuous coverage from TeV to EeV energies, and probe if the astrophysical neutrino spectrum continues unbroken to higher energies or if new classes of accelerators begin to dominate. Excitingly, the radio array will probe for cosmogenic neutrinos with sufficient sensitivity to detect events even if the flux of cosmic ray primaries is very heavy (iron rich). If Gen2-Radio detects no events, it would exclude as astrophysical neutrino accelerators many source populations which trace the star formation rate~\cite{Aartsen:2020fgd}.

%Something about BSM should probably go here
In observing high-energy neutrinos, IceCube-Gen2 will also probe fundamental physics. For example, IceCube-Gen2 will enable the first measurement of the neutrino cross-section at PeV and EeV energies, well beyond the reach of existing particle accelerators. The observatory also serves as a large-volume detector for exotic and beyond the standard model physics programs, e.g. searches for magnetic monopoles or sterile neutrinos~\cite{Aartsen:2020fgd}.

\section{Instrumentation Development}
\label{sec:instrumentation}

\begin{figure}[]
    \centering
    \includegraphics[trim={2cm 4cm 2cm 0},clip, width=4cm]{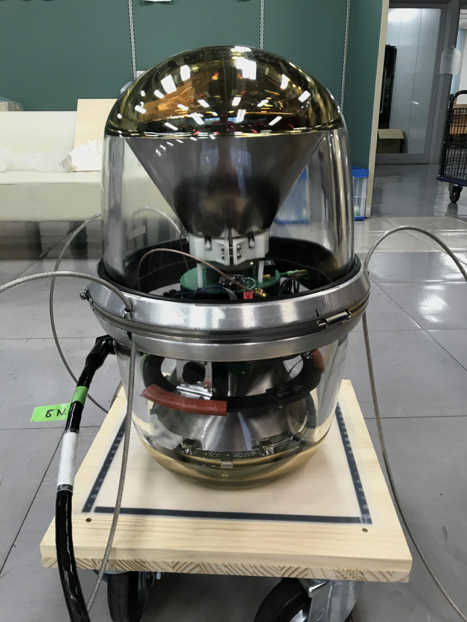}
    \includegraphics[width=4cm]{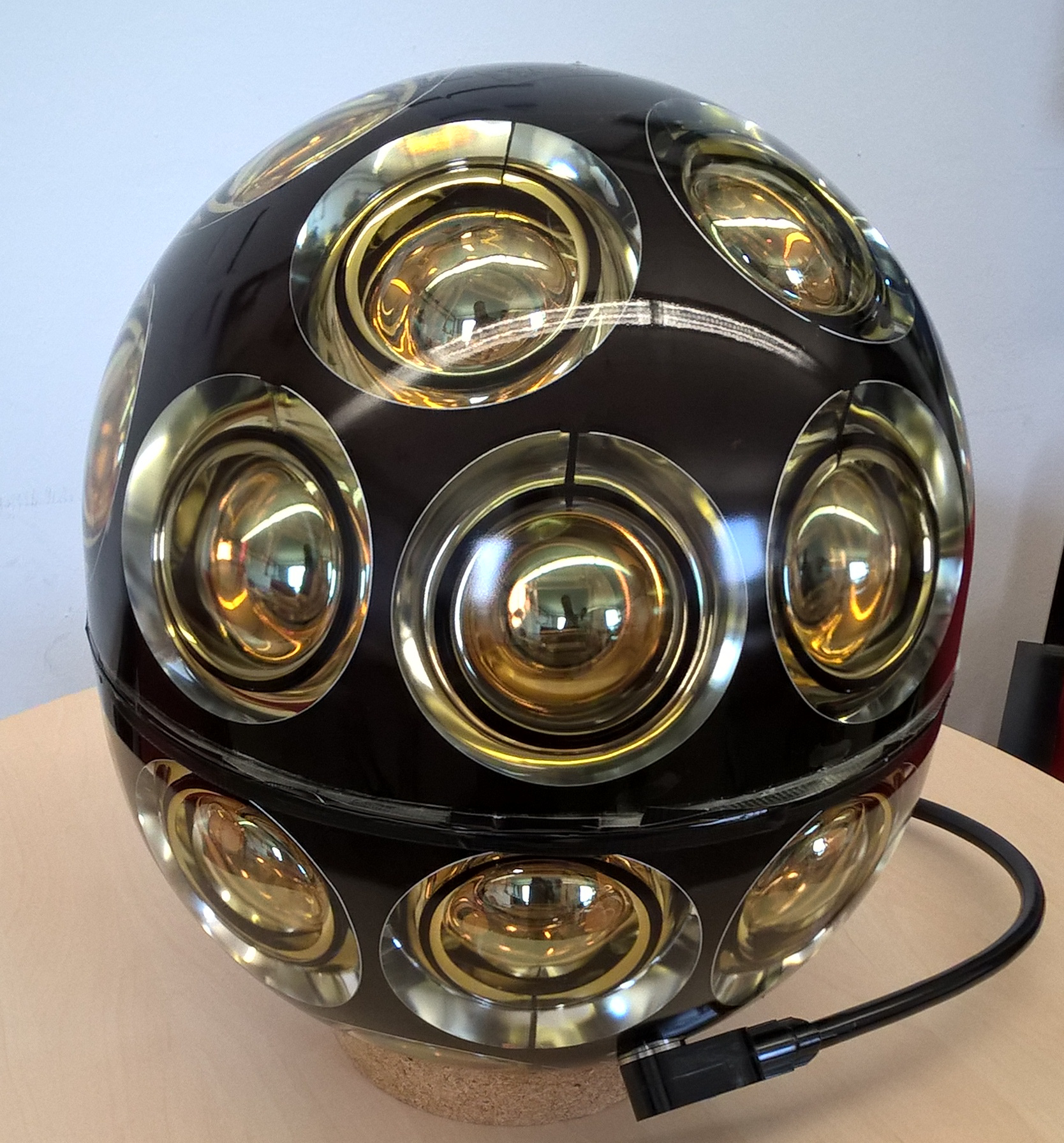}
     \includegraphics[width=4cm]{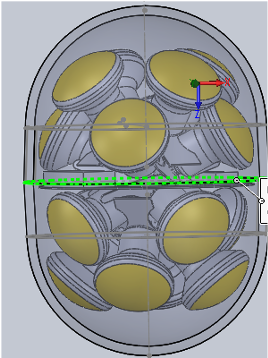}
    \caption{Figures of two pixelated modules being deployed in the IceCube-Upgrade: D-Eggs (left) and mDOMs (middle). At right, a drawing of a potential IceCube-Gen2 OM.}
    \label{fig:optical_modules}
\end{figure}

For IceCube-Gen2 optical, the OMs will be \textit{pixelated}, containing multipile PMTs to achieve triple the photocathode area of a IceCube DOM while also providing directional information per-OM. Such pixelated modules are in use in the KM3NeT telescope, and development of such modules for IceCube has been completed for the IceCube-Upgrade. Fig.~\ref{fig:optical_modules} depicts two such OMs which will be deployed: the D-Egg~\cite{Ishihara:2017vxn} and mDOM~\cite{Classen:2019tlb}. An example conceptual design for a Gen2 OM is also depicted in the figure~\cite{Basu:VLVNT2021}. Other novel photon detectors are also under consideration, and will be tested in the Upgrade; for example wavelength shifting sensors~\cite{Pollmann:VLVNT2021} which would record photons in the ultraviolet where Cherenkov radiation is more intense, and then shift them to lower frequencies to be recorded in PMTs. For the surface array, early models of the scintillator panels and radio antennas planned for the surface arrray have already been developed are being deployed as part of an enhancement to the existing IceTop detector~\cite{Haungs:2019ylq}.

The radio array for IceCube-Gen2 draws extensive heritage from existing experiments. The technology for the deep antennas as well as the phased array triggering system were piloted by RICE~\cite{2012PhRvD..85f2004K} and ARA~\cite{Allison:2019xtn, Allison:2018ynt}, while the deployment of antennas at the surface was piloted by the ARIANNA~\cite{Anker:2019rzo} experiment. Some new technologies are envisioned for IceCube-Gen2, such as LTE-based communications for the transfer of data and the control of station operations. This new technology, along with joint surface and deep components, will be deployed in the RNO-G experiment~\cite{Aguilar:2020xnc} beginning in Summer 2021.

For both the optical array and the radio array, holes must be drilled into the glacier in which to deploy instrumentation. Efforts are underway to make this drilling process more efficient, and to produce higher quality final holes. For the optical array, the holes must be melted $\sim2.6$\,km into the glacier. The drill from IceCube is being modified for enhanced performance. For example, the drill will be moved on large sled modules, and delivered by tractor-drawn traverse instead air to reduce logistical footprint. Additionally, the holes will be ``degassed" in an effort to remove the bubble column which concentrated at the center of IceCube holes and leads to increased light scattering~\cite{Rongen:2019wsh}. For the radio array, a dry-drilling technique is envisioned. Based on the ``RAID" drill of the British Antarctic Survey, it removes the ice as chips and appears capable of reaching 200\,m depth within 20 hours~\cite{BigRAID}.

\section{Conclusion and Outlook}
\label{sec:conclusion}

The IceCube-Gen2 neutrino observatory will be a broadband neutrino detector with unprecendent capabilities.  Much of these enhanced sensitivity is driven by new technologies, including pixelated optical modules, radio arrays, and scintillator panels. All of these technologies are under development. The success of IceCube promises an exciting science impact for Gen2.  With an enlarged optical and surface array and a new ultra-high-energy radio array, IceCube-Gen2 will provide opportunities to improve our understanding of the high-energy universe and will play an integral role in the next decade of multimessenger astrophysics.

\section*{Acknowledgements}
The presenter thanks the US National Science Foundation for support through Award AST-1903885.

\clearpage
\bibliographystyle{JHEP}
\bibliography{references}

\providecommand{\href}[2]{#2}\begingroup\raggedright\begin{thebibliography}{10}

\bibitem{Aartsen:2013jdh}
{\scshape IceCube} collaboration, \emph{{Evidence for High-Energy
  Extraterrestrial Neutrinos at the IceCube Detector}},
  \href{https://doi.org/10.1126/science.1242856}{\emph{Science} {\bfseries 342}
  (2013) 1242856} [\href{https://arxiv.org/abs/1311.5238}{{\ttfamily
  1311.5238}}].

\bibitem{Stettner:2019tok}
{\scshape IceCube} collaboration, \emph{{Measurement of the Diffuse
  Astrophysical Muon-Neutrino Spectrum with Ten Years of IceCube Data}},
  \href{https://doi.org/10.22323/1.358.1017}{\emph{PoS} {\bfseries ICRC2019}
  (2020) 1017} [\href{https://arxiv.org/abs/1908.09551}{{\ttfamily
  1908.09551}}].

\bibitem{Aartsen:2020aqd}
{\scshape IceCube} collaboration, \emph{{Characteristics of the diffuse
  astrophysical electron and tau neutrino flux with six years of IceCube high
  energy cascade data}},
  \href{https://doi.org/10.1103/PhysRevLett.125.121104}{\emph{Phys. Rev. Lett.}
  {\bfseries 125} (2020) 121104}
  [\href{https://arxiv.org/abs/2001.09520}{{\ttfamily 2001.09520}}].

\bibitem{Aartsen:2018vez}
{\scshape IceCube} collaboration, \emph{{Measurements using the inelasticity
  distribution of multi-TeV neutrino interactions in IceCube}},
  \href{https://doi.org/10.1103/PhysRevD.99.032004}{\emph{Phys. Rev. D}
  {\bfseries 99} (2019) 032004}
  [\href{https://arxiv.org/abs/1808.07629}{{\ttfamily 1808.07629}}].

\bibitem{Abbasi:2020jmh}
{\scshape IceCube} collaboration, \emph{{The IceCube high-energy starting event
  sample: Description and flux characterization with 7.5 years of data}},
  \href{https://arxiv.org/abs/2011.03545}{{\ttfamily 2011.03545}}.

\bibitem{IceCube:2021rpz}
{\scshape IceCube} collaboration, \emph{{Detection of a particle shower at the
  Glashow resonance with IceCube}},
  \href{https://doi.org/10.1038/s41586-021-03256-1}{\emph{Nature} {\bfseries
  591} (2021) 220}.

\bibitem{IceCube:2020abv}
{\scshape IceCube} collaboration, \emph{{Measurement of Astrophysical Tau
  Neutrinos in IceCube's High-Energy Starting Events}},
  \href{https://arxiv.org/abs/2011.03561}{{\ttfamily 2011.03561}}.

\bibitem{IceCube:2018cha}
{\scshape IceCube} collaboration, \emph{{Neutrino emission from the direction
  of the blazar TXS 0506+056 prior to the IceCube-170922A alert}},
  \href{https://doi.org/10.1126/science.aat2890}{\emph{Science} {\bfseries 361}
  (2018) 147} [\href{https://arxiv.org/abs/1807.08794}{{\ttfamily
  1807.08794}}].

\bibitem{IceCube:2018dnn}
{\scshape IceCube, Fermi-LAT, MAGIC, AGILE, ASAS-SN, HAWC, H.E.S.S., INTEGRAL,
  Kanata, Kiso, Kapteyn, Liverpool Telescope, Subaru, Swift NuSTAR, VERITAS,
  VLA/17B-403} collaboration, \emph{{Multimessenger observations of a flaring
  blazar coincident with high-energy neutrino IceCube-170922A}},
  \href{https://doi.org/10.1126/science.aat1378}{\emph{Science} {\bfseries 361}
  (2018) eaat1378} [\href{https://arxiv.org/abs/1807.08816}{{\ttfamily
  1807.08816}}].

\bibitem{Blaufuss:2019fgv}
{\scshape IceCube} collaboration, \emph{{The Next Generation of IceCube
  Real-time Neutrino Alerts}},
  \href{https://doi.org/10.22323/1.358.1021}{\emph{PoS} {\bfseries ICRC2019}
  (2020) 1021} [\href{https://arxiv.org/abs/1908.04884}{{\ttfamily
  1908.04884}}].

\bibitem{Stein:2020xhk}
R.~Stein et~al., \emph{{A tidal disruption event coincident with a high-energy
  neutrino}}, \href{https://doi.org/10.1038/s41550-020-01295-8}{\emph{Nature
  Astron.} {\bfseries 5} (2021) 510}
  [\href{https://arxiv.org/abs/2005.05340}{{\ttfamily 2005.05340}}].

\bibitem{Adrian-Martinez:2016fdl}
{\scshape KM3NeT} collaboration, \emph{{Letter of intent for KM3NeT 2.0}},
  \href{https://doi.org/10.1088/0954-3899/43/8/084001}{\emph{J. Phys. G}
  {\bfseries 43} (2016) 084001}
  [\href{https://arxiv.org/abs/1601.07459}{{\ttfamily 1601.07459}}].

\bibitem{1997APh.....7..263B}
I.~A. {Belolaptikov} et~al., \emph{{The Baikal underwater neutrino telescope:
  Design, performance, and first results}},
  \href{https://doi.org/10.1016/S0927-6505(97)00022-4}{\emph{Astroparticle
  Physics} {\bfseries 7} (1997) 263}.

\bibitem{Agostini:2020aar}
{\scshape P-ONE} collaboration, \emph{{The Pacific Ocean Neutrino Experiment}},
  \href{https://doi.org/10.1038/s41550-020-1182-4}{\emph{Nature Astron.}
  {\bfseries 4} (2020) 913} [\href{https://arxiv.org/abs/2005.09493}{{\ttfamily
  2005.09493}}].

\bibitem{Aartsen:2020fgd}
{\scshape IceCube-Gen2} collaboration, \emph{{IceCube-Gen2: the window to the
  extreme Universe}}, \href{https://doi.org/10.1088/1361-6471/abbd48}{\emph{J.
  Phys. G} {\bfseries 48} (2021) 060501}
  [\href{https://arxiv.org/abs/2008.04323}{{\ttfamily 2008.04323}}].

\bibitem{Ishihara:2019aao}
{\scshape IceCube} collaboration, \emph{{The IceCube Upgrade - Design and
  Science Goals}}, \href{https://doi.org/10.22323/1.358.1031}{\emph{PoS}
  {\bfseries ICRC2019} (2021) 1031}
  [\href{https://arxiv.org/abs/1908.09441}{{\ttfamily 1908.09441}}].

\bibitem{Rongen:VLVNT2021}
{\scshape IceCube} collaboration, \emph{{Advances in IceCube ice modelling and
  what to expect from the Upgrade}}, {\emph{JINST} {\bfseries VLVnT21} (these
  proceedings) }.

\bibitem{Ishihara:2017vxn}
{\scshape IceCube-Gen2} collaboration, \emph{{Overview and performance of the
  D-Egg optical sensor for IceCube-Gen2}},
  \href{https://doi.org/10.22323/1.301.1051}{\emph{PoS} {\bfseries ICRC2017}
  (2018) 1051}.

\bibitem{Classen:2019tlb}
{\scshape IceCube} collaboration, \emph{{A multi-PMT Optical Module for the
  IceCube Upgrade}}, \href{https://doi.org/10.22323/1.358.0855}{\emph{PoS}
  {\bfseries ICRC2019} (2020) 855}
  [\href{https://arxiv.org/abs/1908.10802}{{\ttfamily 1908.10802}}].

\bibitem{Basu:VLVNT2021}
{\scshape IceCube} collaboration, \emph{{A long multi-PMT optical sensor for
  IceCube-Gen2}}, {\emph{JINST} {\bfseries VLVnT21} (these proceedings) }.

\bibitem{Pollmann:VLVNT2021}
{\scshape IceCube} collaboration, \emph{{Hybrid optical modules for IceCube
  Extensions}}, {\emph{JINST} {\bfseries VLVnT21} (these proceedings) }.

\bibitem{Haungs:2019ylq}
{\scshape IceCube} collaboration, \emph{{A Scintillator and Radio Enhancement
  of the IceCube Surface Detector Array}},
  \href{https://doi.org/10.1051/epjconf/201921006009}{\emph{EPJ Web Conf.}
  {\bfseries 210} (2019) 06009}
  [\href{https://arxiv.org/abs/1903.04117}{{\ttfamily 1903.04117}}].

\bibitem{2012PhRvD..85f2004K}
I.~{Kravchenko}, S.~{Hussain}, D.~{Seckel}, D.~{Besson}, E.~{Fensholt},
  J.~{Ralston} et~al., \emph{{Updated results from the RICE experiment and
  future prospects for ultra-high energy neutrino detection at the south
  pole}}, \href{https://doi.org/10.1103/PhysRevD.85.062004}{\emph{Phys. Rev. D}
  {\bfseries 85} (2012) 062004}
  [\href{https://arxiv.org/abs/1106.1164}{{\ttfamily 1106.1164}}].

\bibitem{Allison:2019xtn}
{\scshape ARA} collaboration, \emph{{Constraints on the diffuse flux of
  ultrahigh energy neutrinos from four years of Askaryan Radio Array data in
  two stations}},
  \href{https://doi.org/10.1103/PhysRevD.102.043021}{\emph{Phys. Rev. D}
  {\bfseries 102} (2020) 043021}
  [\href{https://arxiv.org/abs/1912.00987}{{\ttfamily 1912.00987}}].

\bibitem{Allison:2018ynt}
{\scshape ARA} collaboration, \emph{{Design and performance of an
  interferometric trigger array for radio detection of high-energy neutrinos}},
  \href{https://doi.org/10.1016/j.nima.2019.01.067}{\emph{Nucl. Instrum. Meth.
  A} {\bfseries 930} (2019) 112}
  [\href{https://arxiv.org/abs/1809.04573}{{\ttfamily 1809.04573}}].

\bibitem{Anker:2019rzo}
{\scshape ARIANNA} collaboration, \emph{{A search for cosmogenic neutrinos with
  the ARIANNA test bed using 4.5 years of data}},
  \href{https://doi.org/10.1088/1475-7516/2020/03/053}{\emph{JCAP} {\bfseries
  03} (2020) 053} [\href{https://arxiv.org/abs/1909.00840}{{\ttfamily
  1909.00840}}].

\bibitem{Aguilar:2020xnc}
{\scshape RNO-G} collaboration, \emph{{Design and Sensitivity of the Radio
  Neutrino Observatory in Greenland (RNO-G)}},
  \href{https://doi.org/10.1088/1748-0221/16/03/P03025}{\emph{JINST} {\bfseries
  16} (2021) P03025} [\href{https://arxiv.org/abs/2010.12279}{{\ttfamily
  2010.12279}}].

\bibitem{Rongen:2019wsh}
M.~Rongen, \emph{{Calibration of the IceCube Neutrino Observatory}},  other
  thesis, 11, 2019,
  \href{https://doi.org/10.18154/RWTH-2019-09941}{10.18154/RWTH-2019-09941},
  [\href{https://arxiv.org/abs/1911.02016}{{\ttfamily 1911.02016}}].

\bibitem{BigRAID}
J.~Rix, ``Bigraid.''
  \url{https://events.icecube.wisc.edu/event/128/contributions/7295/}, 2021.

\end{thebibliography}\endgroup

\end{document}